\documentclass[twocolumn,english,prl,showpacs,preprintnumbers,superscriptaddress]{revtex4-1}
\usepackage[utf8]{inputenc}
\usepackage{amsmath}
\usepackage{amssymb}
\usepackage{graphicx}
\usepackage{textcomp}
\usepackage{bm}
\usepackage[right]{eurosym}
\usepackage{float}
\usepackage[english]{babel}
\usepackage{blindtext}
\usepackage{babel}
\usepackage{lineno}
\usepackage{booktabs}
\usepackage{longtable}

\begin{document}
\author{R. Michiels}
\affiliation{Institute of Physics, University of Freiburg, 79104 Freiburg, Germany}
\email{rupert.michiels@physik.uni-freiburg.de}
\author{A. C. LaForge}
\affiliation{Institute of Physics, University of Freiburg, 79104 Freiburg, Germany}
\affiliation{Department of Physics, University of Connecticut, Storrs, Connecticut, 06269, USA}
\author{M. Bohlen}
\affiliation{Institute of Physics, University of Freiburg, 79104 Freiburg, Germany}
\author{C. Callegari}
\affiliation{Elettra-Sincrotrone Trieste S.C.p.A., 34149 Basovizza, Trieste, Italy}
\author{A. Clark}
\affiliation{Laboratory of Molecular Nanodynamics, Ecole Polytechnique F{\'e}d{\'e}rale de Lausanne, 1015 Lausanne, Switzerland}
\author{A. von Conta}
\affiliation{Laboratorium f{\"u}r Physikalische Chemie, ETH Z\"urich, 8093 Z\"urich, Switzerland}
\author{M. Coreno}
\affiliation{ISM-CNR, Istituto di Struttura della Materia, LD2 Unit, 34149 Trieste, Italy}
\author{M. Di Fraia}
\affiliation{Elettra-Sincrotrone Trieste S.C.p.A., 34149 Basovizza, Trieste, Italy}
\author{M. Drabbels}
\affiliation{Laboratory of Molecular Nanodynamics, Ecole Polytechnique F{\'e}d{\'e}rale de Lausanne, 1015 Lausanne, Switzerland}
\author{P. Finetti}
\affiliation{Elettra-Sincrotrone Trieste S.C.p.A., 34149 Basovizza, Trieste, Italy}
\author{M. Huppert}
\affiliation{Laboratorium f{\"u}r Physikalische Chemie, ETH Z\"urich, 8093 Z\"urich, Switzerland}
\author{V. Oliver}
\affiliation{Laboratory of Molecular Nanodynamics, Ecole Polytechnique F{\'e}d{\'e}rale de Lausanne, 1015 Lausanne, Switzerland}
\author{O. Plekan}
\affiliation{Elettra-Sincrotrone Trieste S.C.p.A., 34149 Basovizza, Trieste, Italy}
\author{K. C. Prince}
\affiliation{Elettra-Sincrotrone Trieste S.C.p.A., 34149 Basovizza, Trieste, Italy}
\author{S. Stranges}
\affiliation{Department of Chemistry and Drug Technologies, University Sapienza, 00185 Rome, Italy, and Tasc IOM-CNR, Basovizza, Trieste, Italy}
\author{H. J. W\"orner}
\affiliation{Laboratorium f{\"u}r Physikalische Chemie, ETH Z\"urich,	 8093 Z\"urich, Switzerland}
\author{F. Stienkemeier}
\affiliation{Institute of Physics, University of Freiburg, 79104 Freiburg, Germany}

\title[An \textsf{achemso} demo]
  {Evolution and ion kinetics of a XUV-induced nanoplasma in ammonia clusters}
\begin{abstract}

High-intensity extreme ultraviolet (XUV) pulses from a free-electron laser can be used to create a nanoplasma in clusters. In Ref.~[Michiels et al. PCCP, 2020; 22: 7828-7834] we investigated the formation of excited states in an XUV-induced nanoplasma in ammonia clusters. In the present article we expand our previous study with a detailed analysis of the nanoplasma evolution and ion kinetics. We use a time-delayed UV laser as probe to ionize excited states of H and H$_2^+$ in the XUV-induced plasma. Employing covariance mapping techniques, we show that the correlated emission of protons plays an important role in the plasma dynamics. The time-dependent kinetic energy of the ions created by the probe laser is measured, revealing the charge neutralization of the cluster happens on a sub-picosecond timescale. Furthermore, we observe ro-vibrationally excited molecular hydrogen ions H$_2^{+*}$ being ejected from the clusters. We rationalize our data through a qualitative model of a finite-size non-thermal plasma.

\end{abstract}
\date{\today}
\maketitle
\section{Introduction} 

Laser-induced nanoplasmas have been an active field of research in recent years and combine high energy physics~\cite{ditmire1999nuclear} with atomic and molecular quantum dynamics on a nanoscale~\cite{fennel2010laser,saalmann2006mechanisms, saalmann2010cluster}. Research has been fueled by the necessity to understand radiation damage and plasma formation in the single-shot-imaging of nanoparticles~\cite{neutze2000potential}. Many intriguing physical processes have been discovered: for example, nanoplasmas have been investigated as sources of high-energy particles~\cite{trivikram2013anomalous,iwan2012explosion,last2001nuclear,hohenberger2005dynamic} and coherent radiation~\cite{tisch1997investigation,bodi2019high}. Nanoplasma research in rare-gas clusters led to the discovery of enhanced absorption by collective quasi-particle resonance of the electrons~\cite{zweiback1999femtosecond}. Many interesting properties of nanoplasmas come from the enhancement of recombination processes due to the large number of confined electrons and positive ions~\cite{schutte2014tracing,schutte2015recombination}. The decay of a nanoplasma is governed by a complex interplay of Coulomb explosion and hydrodynamic forces leading to shock shells in the outer Debye layer~\cite{kaplan2003shock,arbeiter2011rare,schutte2015real,schutte2017tracing}. Single-shot X-ray diffraction imaging of the nanoplasma evolution in rare-gas clusters revealed sub-picosecond dynamics~\cite{gorkhover2016femtosecond}, and a core-shell structure~\cite{rupp2020imaging}.

Nanoplasmas in clusters can be induced using long wavelength radiation (infrared (IR), or near infrared (NIR)) on one hand, or, on the other hand, short-wavelength extreme ultraviolet (XUV) radiation. The two regimes can be distinguished using the Keldysh parameter $\gamma \propto 1/\lambda$, where $\lambda$ is the wavelength of the radiation~\cite{keldysh1964zh,faisal1994strongly,reiss1996coulomb}. A Keldysh parameter of $\gamma \ll 1$ denotes the strong-field regime where photoionization proceeds via tunneling processes~\cite{zheltikov2016keldysh}. On the other hand, the regime where $\gamma \gg 1$ is called the weak-field regime. Photoionization in the weak-field regime proceeds via single- or multi-photon interactions~\cite{zheltikov2016keldysh}. This applies to XUV-induced nanoplasmas~\cite{arbeiter2010ionization}, where as a first approximation, electrons are created with  kinetic energies $E_{\text{ele}} = h\nu-{IP}$. Here, $h\nu$ is the energy of a single XUV photon and $IP$ is the first ionization potential of the species.

Concerning plasma dynamics, molecular clusters with multiple components differ significantly from homogeneous atomic clusters. When hydrogen is among the constituents, the nuclear dynamics speeds up and the lightweight protons offer an efficient pathway for cooling and charge neutralization~\cite{last2004electron,andreev2010quasi,di2013proton}. Nanoplasmas in molecular clusters have been studied previously with tabletop~\cite{snyder1996femtosecond,card1997covariance,niu2005covariance,wang2008cluster} as well as free-electron lasers~\cite{iwan2012explosion,michiels2020time}, showing significant fragmentation of the molecules and the generation of high-energy ions. Calculations predict the inner charge state and temperature in the plasma core to be much lower in (CH$_4$)$_n$ molecular clusters, when compared to atomic C$_n$ clusters~\cite{di2013proton}. Previous pump-probe experiments with nanosecond lasers were unable to resolve the fast plasma dynamics happening on a sub picosecond timescale. In a recent femtosecond XUV-pump UV-probe time-resolved experiment on ammonia clusters, we investigated the dynamics of molecular and atomic states upon nanoplasma formation~\cite{michiels2020time}.

In the present work, we extend these studies with a detailed analysis of the nanoplasma evolution and ion kinetics. First, we will address the kinetic energy of H$^+$ emitted from the nanoplasma and use covariance mapping to analyze how energy is dissipated from the clusters via high-kinetic-energy protons. Secondly, the photoionization of H$^*$ and the photodissociation of H$_2^+$ by the probe laser are discussed. Finally, we analyze time-dependent kinetic energy distributions of H$^+$ and photoelectrons upon UV-ionization of H$^*(n=2)$. We discuss how the observations allow conclusions concerning the lifetime of the Coulomb potential at the cluster surface.

\section{Experimental Setup and Methods}
The experiment was performed at the Low Density Matter (LDM) endstation~\cite{lyamayev2013modular} at the seeded FEL FERMI in Trieste, Italy~\cite{allaria2012highly}. Details on the experimental setup can be found in Ref.~\cite{michiels2020time} and Ref.~\cite{lyamayev2013modular}. A jet of neutral ammonia clusters was created via supersonic expansion using a home-built pulsed nozzle. The mean cluster size was $ \langle N \rangle = 2000$ molecules and the cluster size distribution is assumed to be a broad log-normal distribution~\cite{bobbert2002fragmentation}. The cluster jet was crossed perpendicularly with the XUV laser and a 261\,nm UV laser. The interaction region was in the focus of a combined velocity-map-imaging (VMI) and Wiley-McLaren~\cite{wiley1955time} type ion time-of-flight (ToF) spectrometer. High intensity XUV pulses were used to multiply ionize the ammonia clusters. The FEL pulse intensity in the interaction region was $I_{\text{XUV}}\approx 3\cdot10^{12}$\,W/cm² at 14.3\,eV photon energy, $I_{\text{XUV}} \approx 1\cdot10^{13}$\,W/cm² at 19.2\,eV, 23.8\,eV, 28.6\,eV and 33.4\,eV photon energy and $I_{\text{XUV}}\approx 2\cdot10^{13}$\,W/cm² at 42.9\,eV photon energy. The UV probe pulse had an intensity of $I_{\text{UV}} \approx 2\cdot10^{12}$W/cm². Correlation maps were calculated as Pearson's correlation coefficient~\cite{pearson1895notes,schober2018correlation}. We included partial covariances~\cite{frasinski2016covariance} to compensate for the target density fluctuations and FEL pulse energy fluctuations using the sum intensity of the ion-ToF signal per FEL shot as a control variable. The correlation map was calculated using 30\,000 FEL shots. The shot-to-shot standard deviation of the  FEL pulse energy, as well as the standard deviation of the total ion-ToF sum was $\approx6$\%. We confirm the statistical significance of the correlations with a null hypothesis test using the $p$-value. The $p$-value shows the probability of finding an equal or stronger correlation if the correlation were in fact zero.

\begin{figure}
	\begin{center}
		\includegraphics[width=1\columnwidth]{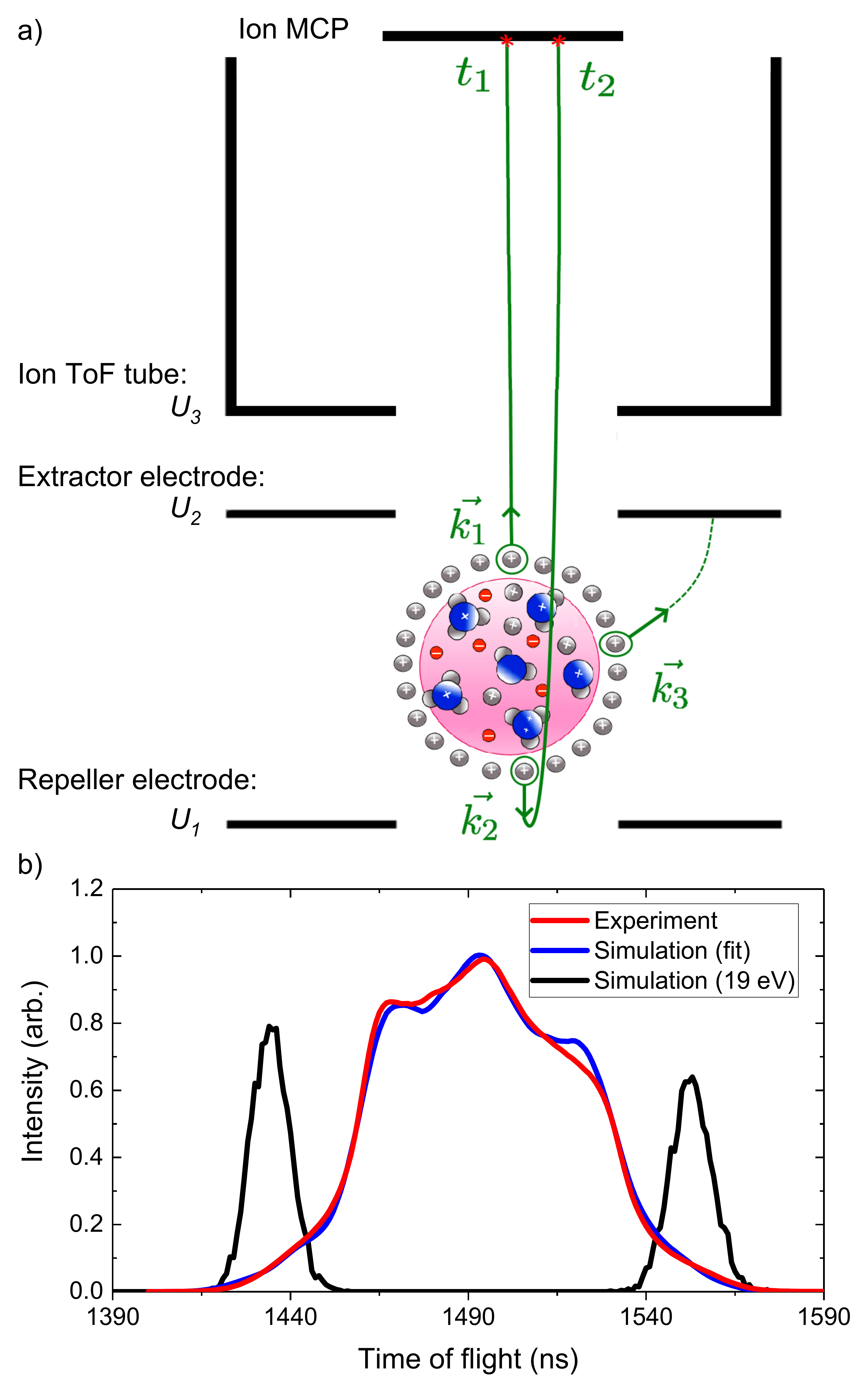}
		\caption{a) Sketch illustrating the ion-ToF geometry and the resulting proton flight paths for three different initial momenta $\vec{k_{1}},\vec{k_{2}},\vec{k_{3}}$. b) Experimental proton-ToF spectrum (red) and best fit from the simulation (blue). The black curve is a simulation for one exact proton kinetic energy (19\,eV), serving as an illustration of the double peak structure.}
		\label{fig:S}
	\end{center}
\end{figure}

In the measured ion-ToF spectrum, the high-kinetic-energy protons create distinct forward and backward peaks corresponding to protons arriving earlier and later than those initially at rest. This process is illustrated in Fig.\,\ref{fig:S}\,a). The flight trajectories of three different protons originating from a highly charged cluster are sketched. The protons created with initial momentum in the forward and backward direction of the spectrometer create separate peaks in the ion-ToF spectrum. Ions with initial velocity components perpendicular to the extraction direction may not be detected. Using ion trajectory simulations and the known geometry of the ToF spectrometer, the initial kinetic energy of the ions can be deduced from their arrival time. An example of the forward and backward peak can be seen in the ToF distribution simulated for protons with 19\,eV kinetic energy, which is shown as the black curve in the Fig.\,\ref{fig:S}\,b).\\

The method used to calculate kinetic energies from the ion-ToF spectra was adapted from Ref.~\cite{bapat2006mean}. We carried out ion-ToF trajectory simulations for discrete integer steps in proton kinetic energy ($E_{\text{H}^+} = 0,1,2,3,...$\,eV) using Simion\textsuperscript{\textregistered}. The experimental spectra were fitted to a linear combination of the simulated spectra using least-squares fitting. An example ToF spectrum (red) and the resulting fit (blue) are shown in Fig.\,\ref{fig:S}\,b). To account for the detector resolution, the simulated ion flight times were convoluted with a Gaussian function. The detector resolution was determined experimentally to be 18\,ns full width at half maximum (FWHM). From the fit parameters, we calculated the mean kinetic energy of the protons. Because the detection efficiency for protons depends on the kinetic energy, we it into account when calculating the mean kinetic energy. The detection efficiency was obtained from the Simion simulations by calculating the portion of protons arriving at the detector. Only about 10\% of isotropically emitted protons with 10\,eV kinetic energy are detected.

\section{Results and Discussion}
\subsection{Correlated emission of protons from the nanoplasma} 
\begin{figure}
	\begin{center}
		\includegraphics[width=1\columnwidth]{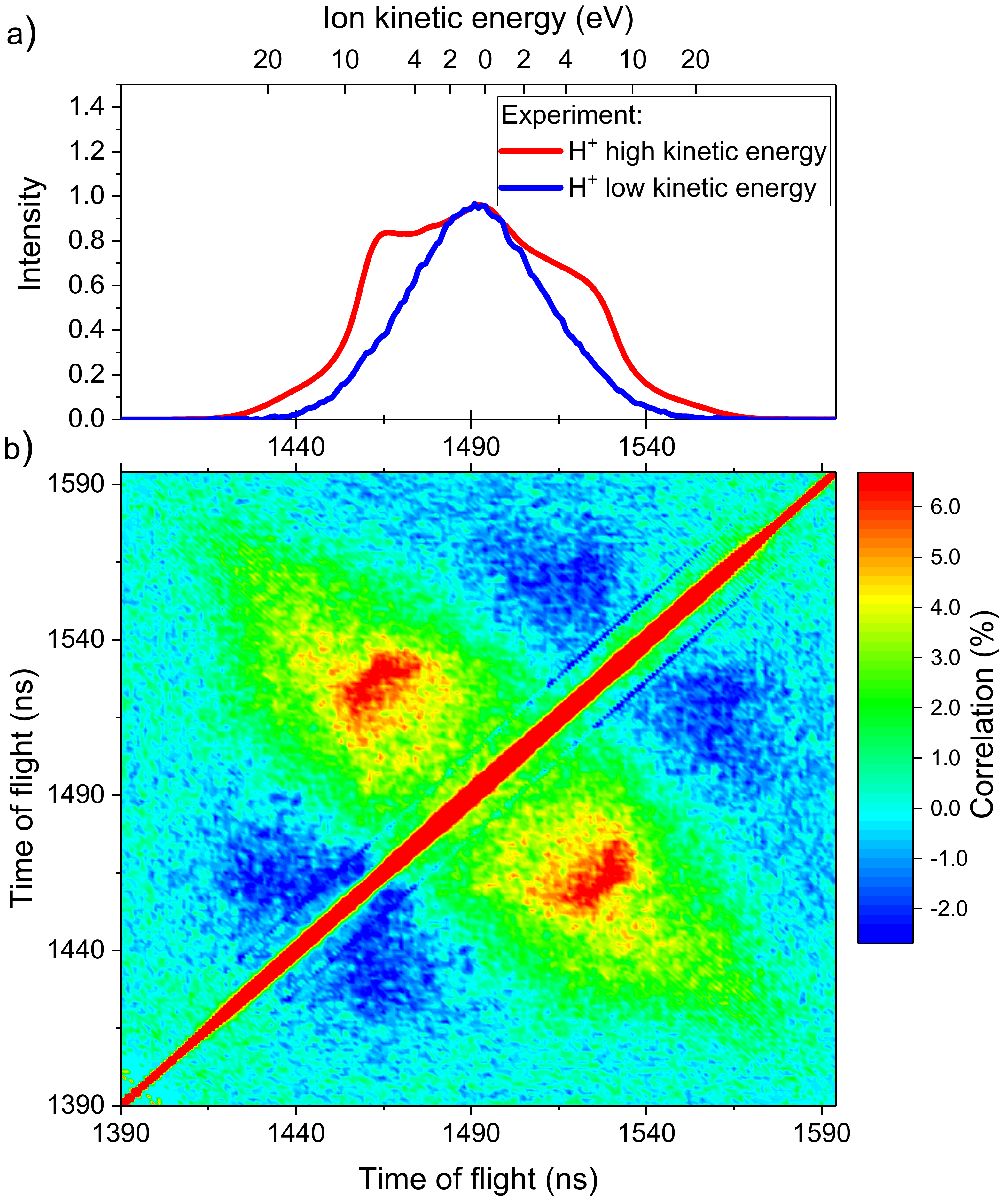}
		\caption{a) Proton peak from the ion-ToF spectrum obtained by irradiation of ammonia clusters with high-intensity XUV pulses ($h\nu = 28.6$\,eV, red) and only the UV pulse (blue). The top axis shows the peak positions obtained from simulations with different proton kinetic energies. b) False-color map of Pearson correlation coefficients of the proton-ToF peak induced by XUV irradiation using 28.6\,eV photon energy.}
		\label{fig:1}
	\end{center}
\end{figure}

A comparison of experimental ToF spectra for protons created with high-intensity XUV pulses (red) and only the UV pulse (blue) is shown in Fig.\,\ref{fig:1}\,a). The red curve is broader than the blue curve, and the shoulders indicate a large contribution from high-kinetic-energy protons. In our experiment, ions are regarded as having a high kinetic energy if they are created with kinetic energies $\geq 4$\,eV. For both curves, the maximum of the intensity is at the center of the peak, which corresponds to the time-of-flight for protons created with negligible kinetic energy. Taking the reduced detection probability for high-kinetic-energy protons into account, we can deduce that the red curve in Fig.\,\ref{fig:1} contains significantly more high-kinetic-energy protons than the blue curve. In the following, we explore the physical process leading to the emission of the high-kinetic-energy protons.

\begin{figure}
	\begin{center}
		\includegraphics[width=1\columnwidth]{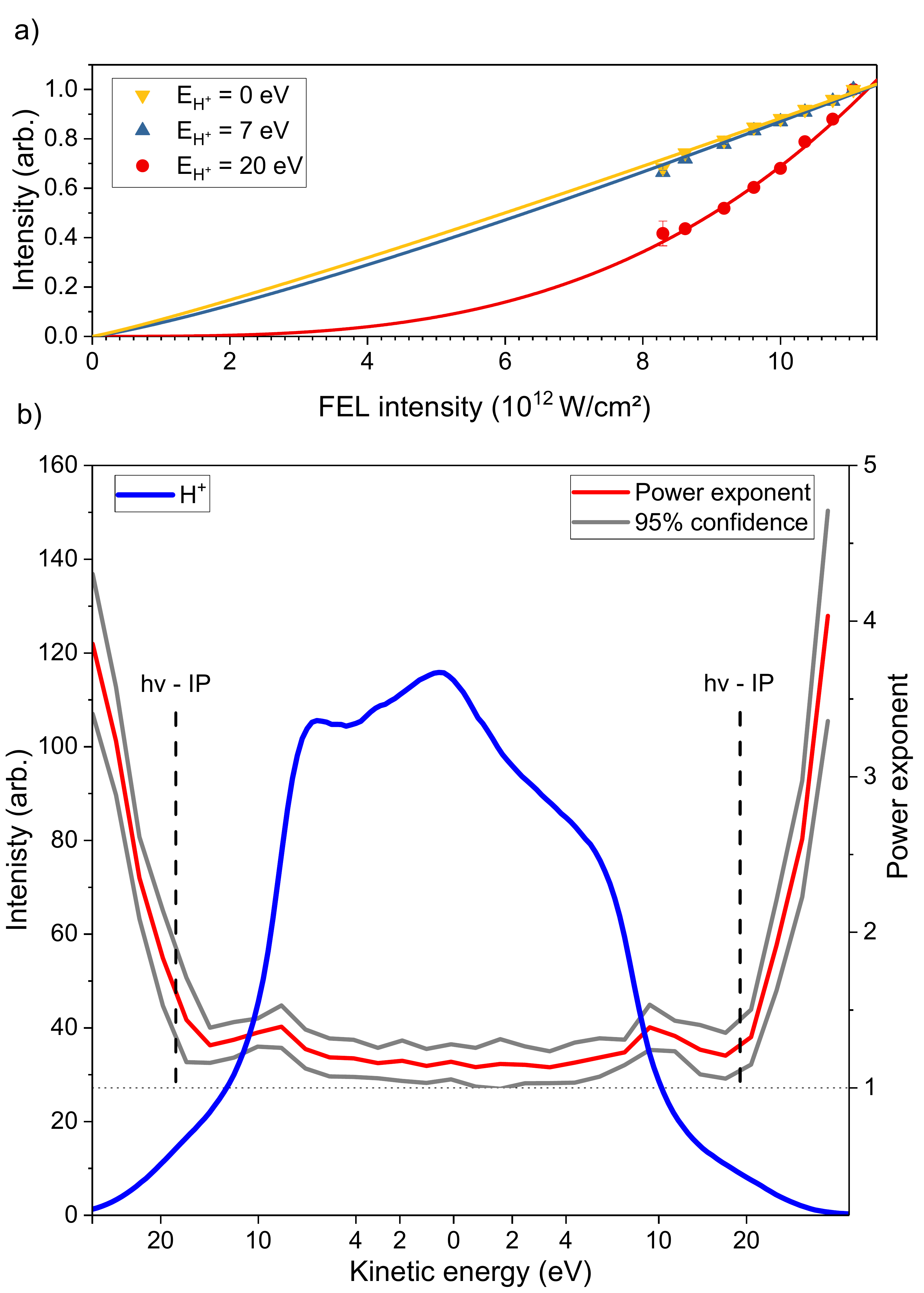}
		\caption{a) Proton signal as a function of FEL intensity for three different proton kinetic energies ($E_{\text{H}^+}$). b) Proton intensity as a function of kinetic energy (blue curve) and the power exponent $k(E_{\text{H}^+})$ (red curve) indicating the pump-power dependence of the proton intensity for different kinetic energies. Data was taken using a pump laser with 28.6\,eV photon energy.} 
		\label{fig:3}
	\end{center}
\end{figure}

The ion ToF-data are recorded for each FEL shot, thus enabling shot-to-shot covariance analysis. Correlations between different ions give information about the underlying fragmentation processes. The correlation map for the proton ToF peak is shown as a false-color plot in Fig.\,\ref{fig:1}\,b). The map shows the correlation between different flight times, with the flight time of one proton on the $x$-axis, and the flight time of the other proton on the $y$-axis. A positive value means that in a given spectrum containing a set of protons arriving at time $t_x$, there is a higher probability to  also detect protons with arrival time $t_y$. A negative value shows a reduced probability, i.e.~that the arrival times $t_x$ and $t_y$ are anti-correlated. On the diagonal line of the map in Fig.\,\ref{fig:1}, the auto-correlation is seen, broadened by the detector resolution. The correlation coefficient is symmetric with respect to the diagonal, we show both sides in order to facilitate projections on the spectrum in the top panel.

In the upper left half of the map, we see a region of strong positive correlations centered at [$t_1$, $t_2$] = [1460\,ns, 1520\,ns], the same flight time as observed for the protons with excess kinetic energy in the top panel. This shows a positive correlation between high-kinetic-energy protons emitted in the forward and backward direction. The positive correlation arises for protons with equal kinetic energy and extends from $E_{\text{H}^+}  \geq 4\text{\,eV}$ to $E_{\text{H}^+} \approx 20\text{\,eV}$. The single-pixel $p$-value of this correlation feature is $p \leq 1\cdot10^{-4}$. Thus, the correlation is significant even for kinetic energies where the collection efficiency is $<10$\%. Additionally, there are spots of negative correlation centered at [$t_1$, $t_2$] = [1520\,ns, 1560\,ns] and [$t_1$, $t_2$] = [1440\,ns, 1470\,ns]. These peaks show an anti-correlation of high-kinetic-energy protons with low-kinetic-energy protons. In the center of the image, the correlation is substantially lower than for the main peaks, converging to zero until overwhelmed by the auto-correlation.

Correlated emission of protons is only possible if there is a common source. To interpret the data, we will look at it in an event-based picture. The correlations observed in the experiment show that there is an underlying event A, which creates high-kinetic-energy protons with equal kinetic energy. Additionally, this event A is negatively correlated to an event B, which creates low kinetic energy protons. Event A needs to be a multi-body fragmentation, either of a doubly charged molecule in the gas phase, or a highly charged cluster. Despite the fact that there are significant amounts of residual non-condensed molecules present in the cluster jet, the correlation observed in Fig.\,\ref{fig:1}\,b) cannot be caused by gas-phase molecules. This is primarily because both detected particles are protons and the parent molecule is ammonia. The only possible fragmentation channel leading to two correlated protons is NH$_3^{2+}$ $\rightarrow$ NH + H$^+$ + H$^+$, a channel which has so far not been observed in photoion-photoion coincidence measurements~\cite{winkoun1986fragmentation, stankiewicz1989double}, or ion-impact dissociation~\cite{bhatt2017formation} on doubly-charged ammonia. Therefore, we assume that event A is associated a single cluster being multiply ionized by absorption of $n \geq 2$ photons.\\

The shape of the positive correlation feature in Fig.\,\ref{fig:1}\,b) resembles a concerted explosion of a multiply charged object~\cite{eland1991dynamics}. This is best explained with a core-shell picture of the highly ionized cluster. The protons in the Debye layer of the cluster are ejected by the Coulomb forces, giving them a radially isotropic momentum distribution. Because the absolute kinetic energy of the protons ejected in one shell is approximately equal, the observed correlation feature follows. The core-shell interpretation can be backed-up by the observation that the kinetic energy of the positively correlated protons is in good agreement with the expected plasma potential $h\nu-IP$~\cite{arbeiter2010ionization}, where $IP$ is the first ionization potential of the ammonia cluster (9.4\,eV~\cite{lindblad2009charge}). In general, charge ejection out of multiply charged clusters happens either as a concerted Coulomb explosion, or as sequential emission of positively charged ions. In a sequential emission process, each cluster gradually cools and creates positive correlation between all kinetic energies of the cooling cascade, including positive correlations between high- and low-kinetic-energy protons. In contrast, we observe a very pronounced correlation between protons of equal kinetic energy, clearly pointing to a concerted ejection of charges. Consequently, the experimental correlations we observe provide strong evidence for a pronounced core-shell nature of the Coulomb explosion of an XUV-induced nanoplasma in ammonia clusters. In this core-shell explosion, a significant amount of charge and energy is taken away by protons with one specific kinetic energy.\\

\begin{figure}
	\begin{center}
		\includegraphics[width=1\columnwidth]{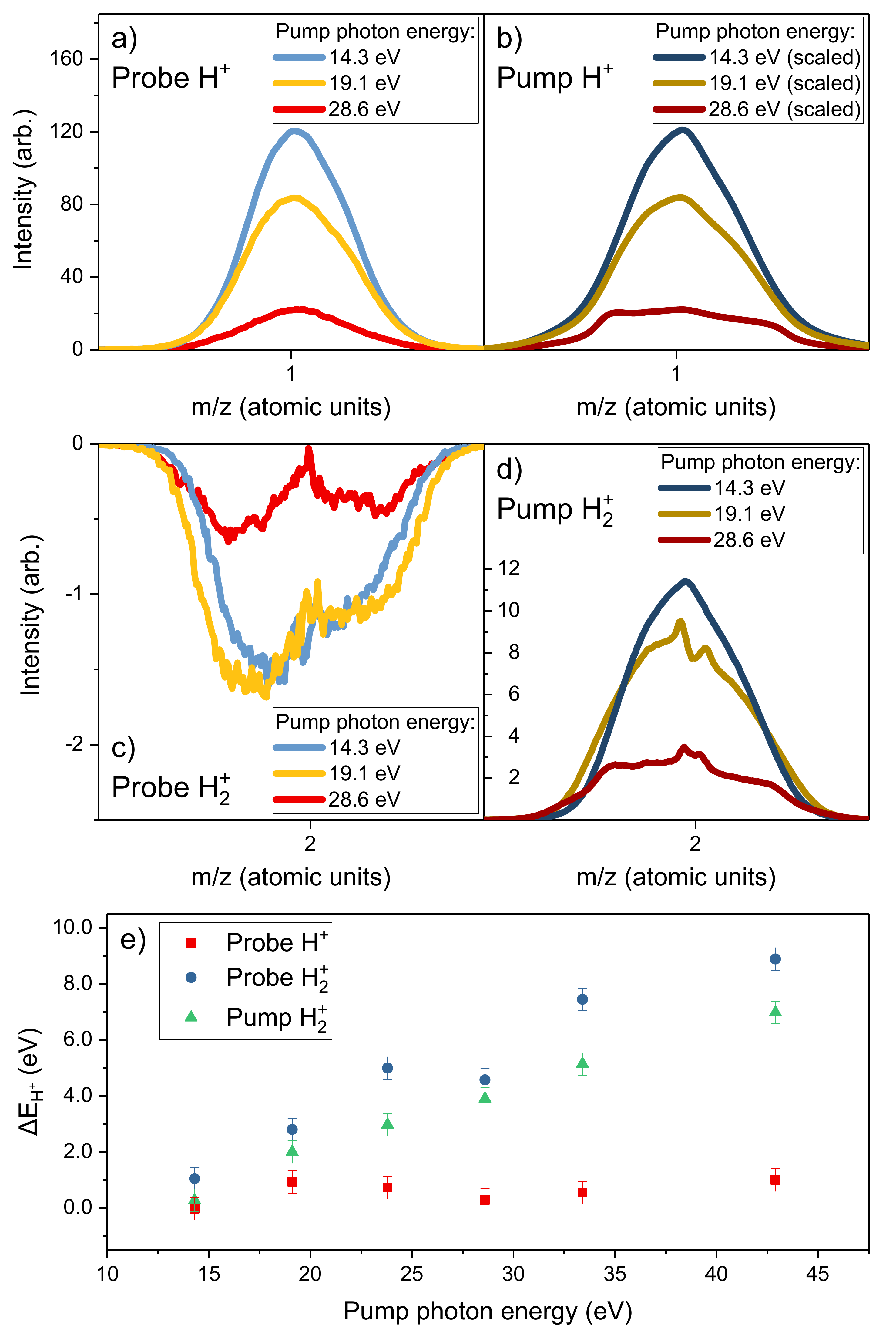}
		\caption{a) Proton-ToF peak created by the probe laser after nanoplasma ignition with different XUV pump wavelengths. b) Corresponding pump proton-ToF peak. c) H$_2^+$-ToF peak created by the probe laser after nanoplasma ignition with different XUV pump wavelengths. d) Corresponding pump H$_2^+$-ToF peak. Note that the small peaks at the center are an artifact from the subtraction of background-gas contributions. e) Mean kinetic energy difference of the H$^+$ and H$_2^+$ ions as a function of the XUV photon energy used for nanoplasma ignition. The values are displayed as difference compared to the mean kinetic energy obtained by UV-ionizing the clusters. This is done in order to distinguish between broadening effects from the cluster environment and kinetic energy release from Coulomb explosion. Red and blue: Probe ions. Green: Pump H$_2^+$ ions. All data shown were obtained at a pump-probe delay of 18\,ps.}
		\label{fig:4}
	\end{center}
\end{figure}

We now take a closer look at how the XUV pulse power influences the kinetic energy of the detected protons. Fig.\,\ref{fig:3}\,b) shows a typical XUV-induced proton ToF peak with strong broadening (blue curve). Additionally, the power exponent $k$ from a power fit: $A(I_{\text{XUV}}) = A_0 I_{\text{XUV}}^k$ is shown for the different areas of the peak (red curve). Here, $A$ is the ion intensity and $I_{\text{XUV}}$ is the XUV pulse intensity. The scaling constant $A_0$ is a free parameter of the fit. Examples of the power fitting for different kinetic energies are given in Fig.\,\ref{fig:3}\,a). Vertical dashed lines in Fig.\,\ref{fig:3}\,b) mark the area where the time of flight corresponds to a proton kinetic energy of $E_{\text{H}^+} = h\nu - IP$. The power coefficient is close to one and approximately constant for all protons with $E_{\text{H}^+} \leq h\nu - IP$. A power coefficient of one shows a linear relation between the number of photons in the XUV pulse and the number of protons detected. We conclude that the number of protons emitted from the nanoplasma rises linearly as a function of XUV pulse intensity. This linear relation is surprising considering the multi-photon nature of the nanoplasma ignition. However, it can be explained with the core-shell structure of the nanoplasma. An increase in XUV pulse intensity cannot increase the plasma potential beyond full frustration. Nonetheless, each additional photon absorbed in the cluster will supply an energy of $h\nu$ to the nanoplasma. A part of this additional energy is dissipated by emitting more protons, and the experimental data show that the relation between the photon flux and the number of protons emitted is linear. Furthermore, we observe that emission of protons with $E_{\text{H}^+} \geq h\nu - IP$ is highly non-linear in XUV pulse intensity. From this we conclude that a large increase in XUV pulse intensity is required to create a plasma potential that is larger than $h\nu - IP$.

\subsection{Probing the nanoplasma with UV laser radiation} 
The previous section discussed the kinetics of protons emitted from highly ionized ammonia clusters. In the following, we will look at the difference in the ion spectrum for pump-probe (XUV+UV pulse) and pump only (XUV pulse). The difference between pump-probe ions and pump-only ions will be called probe ion yield. The ions that are created by the pump alone will be called pump ion yield. The UV laser has a large single-photon ionization cross section for the excited states of molecules and atoms. Absorption of multiple photons is required to ionize the electronic ground states of all involved atoms and molecules. Thus, the effect of the UV probe laser on the neutral molecules is negligibly small when compared to the ionization out of excited states. In the nanoplasma, excited states are created via recombination of free electrons and ions, or by electron-impact excitation~\cite{arbeiter2010ionization,michiels2020time}. 

We will focus on H$^+$ and H$_2^+$ probe ion yields, starting with the asymptotically converged spectrum at a pump-probe delay of 18\,ps. The corresponding probe ion-ToF peaks are shown in Fig.\,\ref{fig:4}\,a)\,and\,c), respectively. We observe that the UV pulse can have two different effects on the ion yield, specifically, producing additional H$^+$ ions, while decreasing the H$_2^+$ ion yield. We observe that the H$_2^+$ ToF spectrum has a double peak structure. In contrast, the H$^+$ ion peak is not significantly broadened. The double peak structure for the H$^+_2$ probe ion yield is particularly pronounced, forming a local minimum in the center, a
feature that is not seen in the pump H$_2^+$ spectrum (see Fig.\,\ref{fig:4}\,d)). The decrease of H$_2^+$ ion yield in the asymptotic difference can only be due to dissociation of H$_2^+$ by the probe laser. We conclude that, despite the fact that H$_2^+$ with low kinetic energy is created by the XUV pump laser, the probe laser primarily dissociates the high-kinetic-energy H$_2^+$.

We will now look at the photodissociation cross section of H$_2^+$ in order to explain how the probe laser selectively dissociates H$_2^+$ with kinetic energy. The dissociation probability in our experiment varies between 10\% at the center of the H$_2^+$ ToF peak, and 50\% on the outer flanks (c.f. Fig.\,\ref{fig:4}\,c)\,and\,d)). To explain a dissociation probability of 10\% with the used probe laser intensity, a photodissociation cross section of $\sigma \approx 0.3$\,Mb is required. Similarly, in order to achieve 50\% dissociation probability, a photodissociation cross section of $\sigma \approx 1.5$\,Mb is required. Previous research has shown that the photodissociation cross section of H$_2^+$ has a strong dependence on the ro-vibrational quantum state of the molecule~\cite{lebedev2003radiative}. The cross sections required for the experimentally observed dissociation probabilities of 10\% and 50\% correspond to a ro-vibrational energy of 2500\,K and 8400\,K, respectively~\cite{lebedev2003radiative}. From this selectivity, we can draw two conclusions: First, a large part of the H$_2^+$ created in the ammonia nanoplasma, particularly the H$_2^+$ with low kinetic energy, has ro-vibrational energies lower than 2500\,K. Secondly, the H$_2^+$ ions that emerge from the nanoplasma with significant kinetic energy do also have a larger ro-vibrational energy.

Quantitative values for the mean kinetic energy of the probe H$_2^+$ and H$^+$ at a pump-probe delay of 18\,ps and the pump H$_2^+$ ions are shown in Fig.\,\ref{fig:4}\,e). The values are displayed as $\Delta E_{\text{H}^+}$, the difference compared to the mean kinetic energy of protons in a spectrum obtained by UV-ionizing the clusters. This is done in order to distinguish between broadening effects from the cluster environment and kinetic energy release from Coulomb explosion. On the $x$-axis, the photon energy of the XUV pump is varied in the range of 14.3\,eV to 42.9\,eV. The kinetic energy of the probe H$_2^+$ ions (blue circles) ranges from 1\,eV (12\,000\,K) to 9\,eV (100\,000\,K), and is larger than the ro-vibrational energy. Furthermore, the mean kinetic energy of probe H$_2^+$ (blue circles) is significantly larger when compared to that of the pump H$_2^+$ (green triangles). The mean kinetic energy of probe and pump H$_2^+$ rises as a function of $h\nu$. The XUV photon energy dependence of the mean kinetic energy of H$_2^+$ shows that a higher XUV photon energy generates a deeper plasma potential. However, the change in mean kinetic energy does not directly correspond to the change in photon energy. This is not surprising, since the total photon flux and the ionization cross section are peaked at 19\,eV. From the difference in the mean kinetic energy of probe and pump H$_2^+$ we conclude that the ro-vibrationally excited H$_2^+$ is primarily created in the Debye layer of highly ionized clusters where the electric field is the strongest. On the other hand, the H$_2^+$ with lower ro-vibrational energy are emitted during the later stages of the nanoplasma evolution.

No significant kinetic energy release can be seen in the probe H$^+$ (red rectangles in Fig.\,\ref{fig:4}\,e)). The primary portion of the probe H$^+$ signal is due to UV ionization out of excited states of atomic hydrogen~\cite{laforge2019real,michiels2020time}. These excited hydrogen atoms are formed according to the reactions~\cite{michiels2020time}:
\begin{equation}
    \text{H}^+ + e^- \rightarrow \text{H}^* \text{\hspace{0.5cm}and}
\end{equation}
\begin{equation}
    \text{NH}_3^{*+} \rightarrow \text{NH}^+_2 + \text{H}^* \text{.}
\end{equation}
From the negligible kinetic energy, we deduce that H$^*$ is not subject to a significant plasma potential at the moment of ionization, i.e., at a pump-probe delay of 18\,ps. High-kinetic-energy protons that are created through the dissociation of H$_2^+$ contribute only a minor part to the probe H$^+$ yield, as can be seen by comparing the absolute scales of Fig.\,\ref{fig:4}\,a)\,and\,c).

\begin{figure}
	\begin{center}
		\includegraphics[width=1\columnwidth]{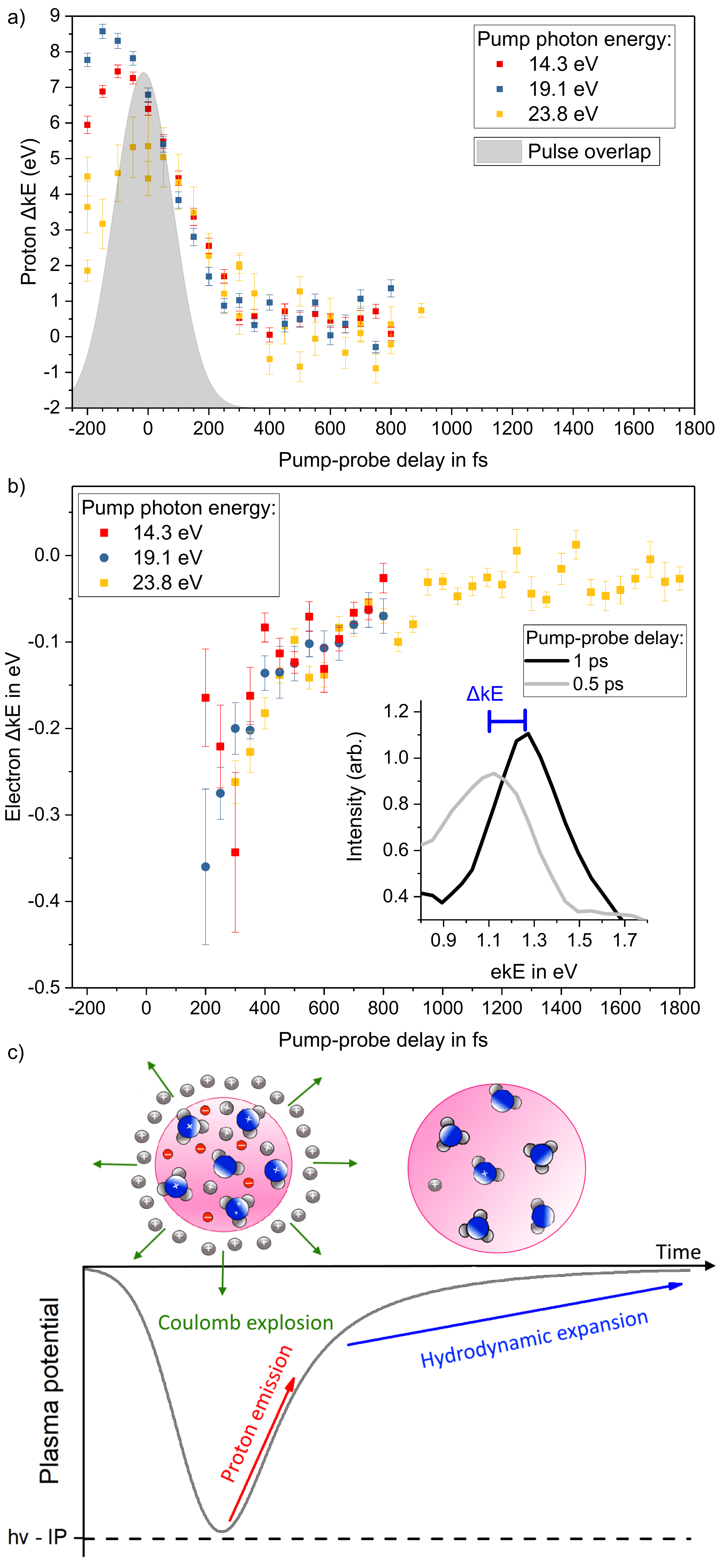}
		\caption{a) Mean kinetic energy of probe-laser-induced H$^+$ versus pump-probe delay for nanoplasma ignition with different XUV photon energies. The shaded area shows the  cross correlation of the two laser pulses deduced from a $1+1'$ ionization of helium atoms. b) Shift in the vertical binding energy of probe laser ionized H$^*$(n\,=\,2) as a function of pump-probe delay. The inset shows an example of the photoelectron peak for two different pump-probe delays. c) Sketch depicting the evolution of the highly ionized ammonia cluster in two phases.}
		\label{fig:5}
	\end{center}
\end{figure}

\subsection{Time-resolved studies on Coulomb explosion}
In the spectra shown in Fig.\,\ref{fig:4}\,a)\,and\,c) we observed that H$^+$ ions created by the probe laser at a pump-probe delay of 18\,ps do not have significant kinetic energy.  For the case that H$^*$ in the cluster is ionized while the nanoplasma is still active, there are two different options. Either the excited hydrogen atom was formed in the bulk of the plasma, in which case the proton will remain inside the cluster; or, the excited hydrogen was formed in the Debye layer of the plasma, in which case the proton will acquire kinetic energy proportional to the plasma potential at the time of ionization. This allows us to use the time-dependent kinetic energy of the probe protons as a probe for the plasma potential. 
Fig.\,\ref{fig:5}\,a) shows the mean kinetic energy of the probe H$^+$ as a function of pump-probe delay. The kinetic energy has a maximum in the vicinity of $t_0$. Afterwards, it decays rapidly and converges at 400\,fs pump-probe delay. The decay time-constant is universal for all pump photon energies and only marginally larger than the temporal pulse overlap of the two laser pulses (depicted as the grey shaded area in Fig.\,\ref{fig:5}\,a)). Using our ToF spectrometer, the kinetic energy of the protons could only be determined for values larger than 500\,meV.\\

Complementary to the ions, we can detect photoelectrons from the probe ionization of H$^*$. The most abundantly populated H$^*$ state is the $n=2$ state~\cite{michiels2020time}, yielding electrons with energy $E_{\text{ele}}$ = 1.35\,eV when ionized with the 4.75\,eV UV photons. If there is a Coulomb potential present, we observe a shift in the kinetic energy of the photoelectrons. The pump-probe delay dependent shift in the vertical binding energy (VBE) of the electrons ($\Delta E_{\text{ele}}$) is displayed in Fig.\,\ref{fig:5}\,b). The inset shows an example of the different peak positions at pump-probe delays of 0.5\,ps and 1\,ps, respectively. A determination of the peak position was only possible for $\Delta E_{\text{ele}} < -400$\,meV. For larger shifts, the peak is strongly broadened and could not be distinguished from the overlapping low energy electrons emitted by the nanoplasma. The VBE shift at a pump-probe delay of 300\,fs is roughly -300\,meV and converges to zero with a half-lifetime of ($200\pm 30$)\,fs.\\

We conclude that the fast charge-equalization seen in the proton kinetic energy is governed by the high-energy Coulomb explosion in the Debye layer. After the Coulomb explosion, the clusters are only mildly charged and hydrodynamic forces are dominant, explaining the slower decay of the remaining -300\,meV plasma potential. At a pump-probe delay of 1\,ps, the plasma potential is completely neutralized. The two-phase expansion of the cluster is illustrated in Fig.\,\ref{fig:5}\,d).

\section{Conclusion}
We induced a nanoplasma in ammonia clusters using high power XUV radiation from the FERMI free-electron laser. Using simultaneous photoelecton and ion detection we have shown that emission of high-kinetic-energy ions plays a significant role in the evolution of XUV-induced nanoplasmas. Using shot-to-shot covariance mapping, we show that protons with kinetic energy $4\text{\,eV} \leq E_{\text{H}^+} \leq 20\text{\,eV}$ are emitted from the nanoplasma in a correlated way. Furthermore, we use a delayed UV laser as probe to ionize excited states of H and H$_2^+$ in the plasma. From the pump-probe dependent mean kinetic energy of the probe-laser induced H$^+$ ions, we get information on the lifetime of the plasma confining potential. We found that the nanoplasma decays in a two-stage process. In the first stage, the potential reduces drastically through the concerted emission of protons. These observations show that the highly-ionized ammonia cluster acts as a core-shell system where the Debye layer of the nanoplasma dissipates a large fraction of the energy contained in the nanoplasma through a concerted Coulomb explosion of protons and other light, positively-charged ions. After the Coulomb explosion, the cluster expansion slows down and hydrodynamic forces become dominant. Using UV-laser-induced dissociation of H$_2^+$, we show that the ion temperatures in the core of the XUV-induced nanoplasma are less than 2500\,K.  

\section*{Acknowledgements}
Funding from the Deutsche Forschungsgemeinschaft (STI 125/19-2, GRK 2079) Carl-Zeiss-Stiftung, grants 200021\_146598 and 200020\_162434 from the Swiss National Science Foundation, as well as the Department of Excellence, Department of Chemistry and Technologies of drugs, and Progetto Ateneo-2016 (prot. n. RG116154C8E02882) of Sapienza University are gratefully acknowledged.

\bibliographystyle{rsc}
\bibliography{wata1}

\end{document}